\documentclass[draft]{agujournal2019}
\usepackage[a4paper]{geometry}
\usepackage{url}
\usepackage{lineno}
\usepackage[inline]{trackchanges}
\usepackage{soul}
\usepackage{array}
\usepackage{tabularx}
\usepackage{graphicx}
\usepackage{amsmath}
\usepackage{amssymb}
\usepackage{booktabs}
\usepackage{hyperref}
\usepackage{multirow}
\usepackage{float}
\usepackage{lmodern}

\draftfalse

\graphicspath{{../}}   

\journalname{Journal of Advances in Modeling Earth Systems (JAMES)}

\begin{document}

\title{Steering Tropical Cyclones Using Small Perturbations in an AI Weather Model}

\authors{
Qin Huang\affil{1},
Moyan Liu\affil{1},
Yeongbin Kwon\affil{2},
Upmanu Lall\affil{1,3}
}

\affiliation{1}{School of Complex Adaptive Systems and Water Institute, Arizona State University, Tempe, AZ, USA}
\affiliation{2}{Department of Civil and Environmental Engineering, Seoul National University, Seoul, Republic of Korea}
\affiliation{3}{Department of Earth and Environmental Engineering and Columbia Water Center, Columbia University, New York, NY, USA}

\correspondingauthor{Qin Huang}{qhuang62@asu.edu}

\begin{keypoints}
\item Hurricane Sandy's track can be substantially altered in an AI weather model through carefully targeted atmospheric perturbations.
\item Local Lagrangian sensitivity and remote Rossby-wave propagation represent two distinct pathways for steering-flow control.
\item The results establish a theoretical sensitivity framework rather than a practical weather-modification strategy.
\end{keypoints}

\begin{abstract}

Tropical cyclone (TC) trajectories are governed by large-scale steering flows and exhibit sensitive dependence on atmospheric initial conditions. Using Hurricane Sandy (2012) in the Aurora AI weather model, we investigate whether targeted thermodynamic perturbations can induce meaningful track deviations. Two distinct perturbation regimes emerge. In the Caribbean, forward finite-time Lyapunov exponent (FTLE) diagnostics identify dynamically sensitive regions within Sandy's steering flow, where perturbations produce substantially larger responses than random placement. In the Pacific, a preferred corridor near $165^\circ$W influences Sandy through Rossby-wave teleconnections, confirmed using Takaya--Nakamura wave activity flux diagnostics. Despite their different physical pathways, both regimes share a common amplification mechanism: small initial perturbations generate modest trajectory offsets that are rapidly amplified when Sandy enters the highly sensitive recurvature region. The largest experiments produce track deviations exceeding 500 km after seven days. These results provide a proof-of-concept demonstration of the Weather Jiu-Jitsu framework, illustrating how targeted perturbations can be amplified through atmospheric dynamics in an AI weather model. Because the required perturbations exceed current operational cloud-seeding capabilities, the experiments should be interpreted as a theoretical sensitivity analysis rather than an operational weather-modification strategy.

\end{abstract}

\section{Introduction}

Tropical cyclones (TCs) cause disproportionate societal impacts relative to their frequency, with track errors of even 100--200~km often determining whether a storm makes catastrophic landfall or remains offshore. Hurricane Sandy (2012) exemplifies this sensitivity: a complex interaction between Sandy's tropical circulation, a deepening mid-latitude trough, and a Greenland blocking anticyclone produced an unusual westward turn toward New Jersey, causing \$65 billion in damage and 233 deaths \citep{Blake2013,Galarneau2013}. This sensitivity raises a fundamental question: can targeted perturbations of the large-scale atmospheric circulation influence TC trajectories?

Early efforts to modify TCs, such as Project Stormfury, focused on disrupting inner-core dynamics through cloud seeding \citep{Willoughby1985}, but these approaches proved ineffective due to the rapid dynamical adjustment of the vortex. An alternative strategy is to target the environmental steering flow, which governs TC motion at synoptic scales \citep{GeorgeGray1976,ChanGray1982}. This perspective is grounded in the theory of sensitive dependence on initial conditions: small perturbations introduced at dynamically sensitive locations can grow nonlinearly and influence large-scale flow evolution \citep{Lorenz1963,Lorenz1969}, with sensitivity concentrated at particular positions within the evolving flow. Rather than opposing atmospheric dynamics, this approach exploits sensitivity at critical junctures, particularly the mid-latitude recurvature zone where TC tracks bifurcate sharply, a concept we refer to as Weather Jiu-Jitsu \citep{Huang2026PLOSWater}. Recent work in low-order dynamical systems demonstrates that perturbations applied during positive-Lyapunov regimes can induce controlled trajectory deviations with minimal energy input \citep{Liu2025_EGU,Liu2026_Lorenz}.

In the full atmospheric system, finite-time Lyapunov exponents (FTLEs) provide a practical diagnostic of flow structure by identifying Lagrangian coherent structures (LCS) that organize transport and delineate dynamical boundaries between flow regimes \citep{HallerYuan2000,Shadden2005,Haller2015,GaraboaPaz2015}. FTLE ridges indicate regions where initially close particle trajectories diverge fastest, marking the boundaries of distinct transport pathways in the steering flow. This suggests a testable hypothesis: thermodynamic perturbations placed at FTLE-identified locations within the TC steering environment should produce larger and more systematic track deviations than perturbations applied elsewhere. A parallel demonstration of this principle is provided by \citet{Liu2026_AR}, who applied FTLE-guided perturbations in the Pacific to reduce integrated water vapor transport (IVT) and shift atmospheric river (AR) landfalls on the California coast.

A TC perturbation pathway requires a different physical framework. TC motion is governed by environmental steering flow that can be teleconnected to upstream sources via Rossby wave propagation \citep{GeorgeGray1976,ChanGray1982}. A thermal perturbation over the Pacific generates anomalous upper-level divergence that acts as a Rossby wave source \citep{Sardeshmukh1988}, exciting a wave packet that propagates eastward along the jet waveguide at group velocity $c_g \approx 25$~m~s$^{-1}$ \citep{HoskinsKaroly1981,HoskinsAmbrizzi1993}. The stationary wavenumber theory of \citet{HoskinsAmbrizzi1993} predicts that certain Pacific longitudes act as preferred wave sources for downstream Atlantic propagation, consistent with the observed sensitivity to longitude in our grid survey. Sandy's westward recurvature involved extratropical transition \citep{Jones2003}, a process in which recurving TCs interact with mid-latitude troughs to produce strongly nonlinear track evolution \citep{Archambault2013}. The recurvature gate is therefore not only spatially specific but also temporally bounded: a Rossby wave packet from the Pacific must arrive at Sandy's position before the storm crosses the trough, constraining which upstream longitudes are effective.

AI-based weather forecast models make such sensitivity experiments computationally tractable. Recent models such as Pangu-Weather, GraphCast, GenCast, and Aurora demonstrate skillful medium-range forecasts \citep{Bi2023,Lam2023,Price2025,Bodnar2025} and enable rapid perturbation experiments. Emerging work confirms that AI models exhibit systematic sensitivity to imposed forcing \citep{Hakim2026,Zhou2026,Peng2026,Liu2026_AR}. Here we use Aurora, which has demonstrated TC track forecast skill superior to other AI models \citep{Bodnar2025,Huang2026_Aurora}, to investigate both the Caribbean FTLE-guided and Pacific Rossby wave perturbation regimes.

We address four questions: (1)~Do FTLE-guided thermodynamic perturbations produce substantial TC track deviations? (2)~Does FTLE-based site selection provide a systematic advantage over random placement, and what spatial and vertical constraints govern this advantage? (3)~Can Pacific upstream perturbations teleconnect to Sandy's recurvature zone via Rossby waves, and what diagnostic identifies productive sites? (4)~What are the key differences between the Caribbean Lagrangian regime and the Pacific wave-mediated regime?

\section{Methods}

\subsection{Aurora AI Weather Forecast Model}

We use the Aurora 0.25$^\circ$ deterministic model \citep{Bodnar2025}, a 1.3-billion-parameter 3D Swin Transformer trained on ERA5 and related Earth system datasets. The model operates on a 721$\times$1440 global grid with 13 pressure levels (1000--50~hPa) and a 6-hour timestep, initialized from two consecutive ERA5 states \citep{ECMWF2019}. Atmospheric state variables include zonal and meridional wind ($u$, $v$), temperature ($T$), specific humidity ($q$), and geopotential ($Z$) at each pressure level. Thermodynamic perturbations are injected by modifying $T$ and $q$ in the initial batch at the designated timestep; Aurora then propagates the perturbed state autoregressively through 168~hours without further intervention.

As a data-driven forecasting system, Aurora does not explicitly enforce physical conservation laws. Perturbation responses are interpreted as sensitivity experiments within a learned dynamical system rather than strictly physical atmospheric trajectories. Its demonstrated short-range skill for TC track prediction \citep{Bodnar2025,Sahu2025,Huang2026_Aurora} justifies its use for probing steering-flow sensitivity at 7-day lead times. All experiments are initialized at 2012-10-23 00:00~UTC, with Hurricane Sandy at 12.6$^\circ$N, 78.4$^\circ$W. TC center positions are extracted using Aurora's built-in MSLP-minimum tracker, which searches a $\pm5^\circ$ box around the prior-step position with 700~hPa geopotential fallback and land masking.

\subsection{Finite-Time Lyapunov Exponent Calculation}

\begin{figure}[htbp]
\noindent\includegraphics[width=\textwidth]{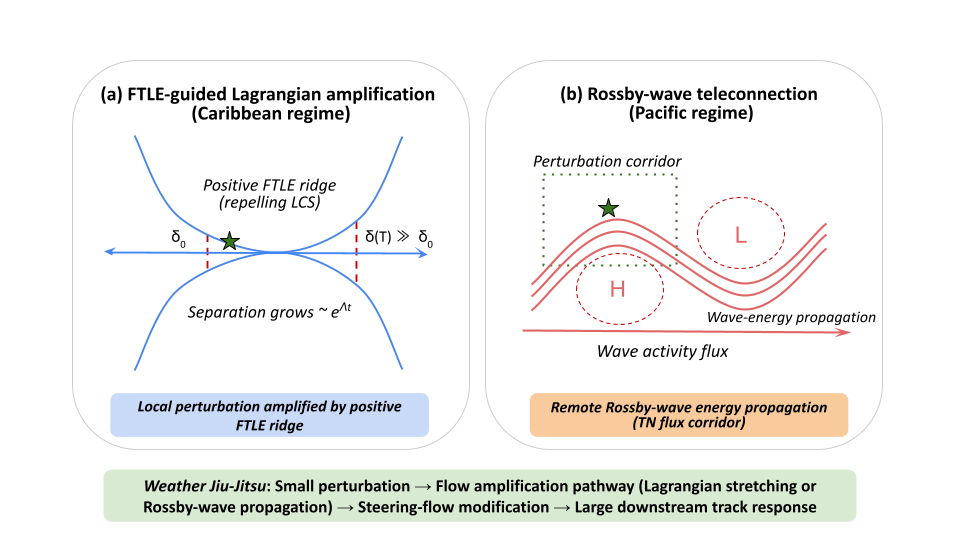}
\caption{Schematic of the two perturbation regimes. (a)~\textit{FTLE-guided Lagrangian amplification (Caribbean regime):} a perturbation (green star, possible seeding site) placed on a positive FTLE ridge ($\Lambda > 0$, repelling LCS) causes an initially small parcel separation $\delta_0$ to grow as $\sim e^{\Lambda t}$, yielding $\delta(T) \gg \delta_0$. The repelling ridge delineates boundaries between distinct transport pathways in Sandy's 500~hPa steering flow; all three seeding sites have $\Lambda = +$1.47 to $+$1.62~day$^{-1}$. (b)~\textit{Rossby-wave teleconnection (Pacific regime):} a perturbation (green star) within the productive 165$^\circ$W corridor excites a Rossby wave packet that propagates eastward along the jet waveguide, diagnosed by Takaya-Nakamura (TN) wave activity flux; the wave energy reaches Sandy's recurvature zone in $\approx$90~h. Bottom: the Weather Jiu-Jitsu strategic chain shared by both regimes: small perturbation $\to$ flow amplification pathway (Lagrangian stretching or Rossby-wave propagation) $\to$ steering-flow modification $\to$ large downstream track response.}
\label{fig:schematic}
\end{figure}

For a flow map $\boldsymbol{\Phi}(\mathbf{x}_0, t_0, \tau)$ advecting a parcel from $\mathbf{x}_0$ at $t_0$ to its position at $t_0 + \tau$, the FTLE is:

\begin{linenomath*}
\begin{equation}
  \Lambda(\mathbf{x}_0, t_0, \tau)
    = \frac{1}{|\tau|}\ln\sqrt{\lambda_{\max}
      \left[\mathbf{C}(\mathbf{x}_0, t_0, \tau)\right]},
  \label{eq:ftle}
\end{equation}
\end{linenomath*}

\noindent where $\mathbf{C} = (\nabla\boldsymbol{\Phi})^\top \mathbf{G} (\nabla\boldsymbol{\Phi})$ is the right Cauchy--Green deformation tensor with spherical metric $\mathbf{G} = \mathrm{diag}(1,\cos^2\!\varphi)$ (where $\varphi$ is latitude), and $\lambda_{\max}$ is its largest eigenvalue \citep{Haller2015,Shadden2005}. The flow-map gradient is computed using dimensionless deg/deg centred finite differences, consistent with \citet{Liu2026_AR}. FTLE ridges are approximate indicators of Lagrangian transport structure, though they do not necessarily correspond to rigorous variational LCS in all flows \citep{Haller2011,HallerSapsis2011}.

We compute \textbf{forward} FTLEs ($\tau = +48$~h) using 4th-order Runge-Kutta (RK4) particle advection with bilinear interpolation from a frozen ERA5 wind snapshot. The Caribbean domain spans 0--35$^\circ$N, 260--300$^\circ$E at 500~hPa ($\sim$22{,}400 particles at 0.25$^\circ$); the Pacific domain spans 25--65$^\circ$N, 130--250$^\circ$E. The 500~hPa level is the canonical TC steering level \citep{GeorgeGray1976}, lying above the TC warm core and below the upper-level outflow. In the Caribbean domain on 2012-10-23, FTLE values range from $-$5.51 to $+$1.92~day$^{-1}$ ($\approx$89\% positive; 85th-percentile threshold $+$0.79~day$^{-1}$).

\subsection{Perturbation Site Selection}

\begin{figure}[htbp]
\noindent\includegraphics[width=\textwidth]{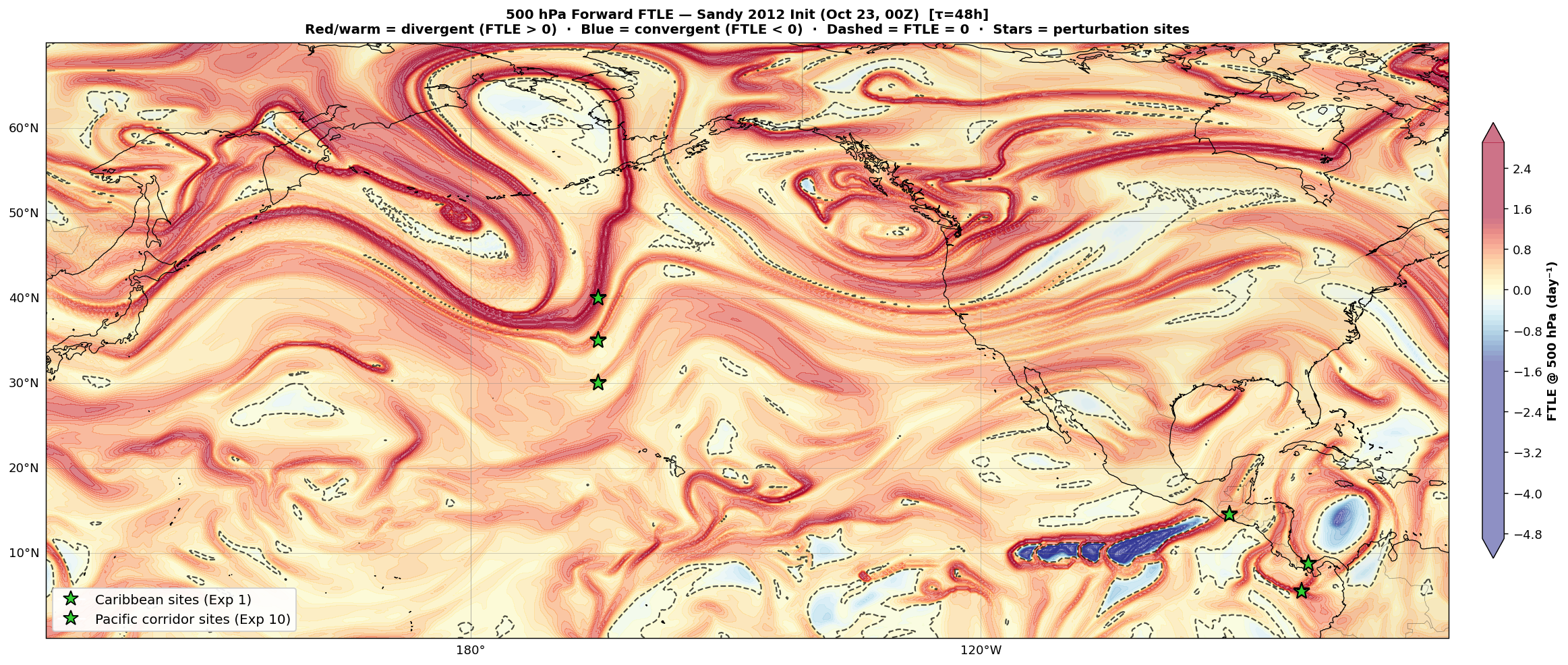}
\caption{48-hour forward FTLE at 500~hPa (2012-10-23 00Z) shown for both perturbation domains on a single combined panel. Positive values (warm colors) indicate repelling Lagrangian boundaries; green star markers show the three FTLE-selected Caribbean seeding sites (Sites~1--3; see Table~\ref{tab:sites}) and the Pacific corridor sites (Exp~10). Strong positive FTLE ridges dominate the Caribbean (89\% positive cells) and the Pacific jet stream. Despite high FTLE coverage in the Pacific, FTLE-selected sites produce only $\sim$99.9~km because the operative constraint is Rossby wave travel-time geometry, not local flow divergence; the productive 165$^\circ$W corridor is identified by the grid survey rather than the FTLE field.}
\label{fig:ftle_field}
\end{figure}

\textbf{Caribbean FTLE regime.} From the 85th-percentile FTLE candidates, sites are selected using two criteria: (1)~spatial clustering via iterative masking within 300~km to ensure distributed locations, and (2)~a distance constraint of 500--1500~km from the TC center. For Sandy, three sites survive both filters (Figure~\ref{fig:ftle_field}; Table~\ref{tab:sites}); each site defines a 300~km circular perturbation region (Exp 1).

\textbf{Pacific wave regime.} Site selection uses a 25-point grid survey (30--50$^\circ$N, 140--250$^\circ$E; Section~\ref{sec:pacific_survey}), with productive sites identified by Rossby wave travel-time reasoning rather than local FTLE ranking. The corridor at 165$^\circ$W (195$^\circ$E) is identified as the productive region after the grid survey (Exp 8); all three corridor sites (30$^\circ$N, 35$^\circ$N, 40$^\circ$N/195$^\circ$E) are used simultaneously in the final corridor experiment (Exp 10).

\subsection{Idealized Thermodynamic Perturbations}

Two perturbation types are used. \textbf{Type A (cloud-seeding physics)} applies coupled temperature--humidity changes motivated by silver iodide nucleation and ice deposition, following \citet{Liu2026_AR}:

\begin{linenomath*}
\begin{equation}
  q_{\text{frozen}} = \eta \cdot q \cdot \min\!\left(1,\,\frac{\text{RH}}{100}\right),
  \quad
  \Delta T = \frac{L_d \, q_{\text{frozen}}}{c_p},
  \quad
  \Delta q = -(1 - f_{\text{fall}}) \, q_{\text{frozen}},
  \label{eq:seeding}
\end{equation}
\end{linenomath*}

\noindent where $\eta = 0.60$ is the freeze efficiency, $L_d = 2.835 \times 10^6$~J~kg$^{-1}$ is the latent heat of deposition, $c_p = 1004$~J~kg$^{-1}$~K$^{-1}$, and $f_{\text{fall}} = 0.80$ is the precipitation fallout fraction. Only grid cells with RH~$>$~75\% within the seeding region are modified. Canonical configuration (Exp~1): $\eta = 60\%$, $r = 300$~km, levels 925/850/700~hPa. \textbf{Type B (Gaussian thermal anomaly)} applies a spatially smooth temperature increment used in the Pacific grid survey:

\begin{linenomath*}
\begin{equation}
  \Delta T(\mathbf{x}) = \Delta T_0 \,\exp\!\left(-\frac{\Delta\mathrm{lon}^2 + \Delta\mathrm{lat}^2}{2\sigma^2}\right),
  \label{eq:gaussian}
\end{equation}
\end{linenomath*}

\noindent with $\Delta T_0 = 2.49$~K, $\sigma = 8^\circ$ ($\approx$890~km), applied to 700/850/925~hPa. This synoptic-scale footprint is designed to probe Rossby wave excitation over the Pacific rather than localized moisture modification. In all experiments, perturbations are applied once and propagated autoregressively.

\subsection{Takaya-Nakamura Wave Activity Flux Diagnostic}

To diagnose Rossby wave propagation from Pacific perturbation sites to Sandy's recurvature zone, we compute the Takaya-Nakamura (TN) wave activity flux at 250~hPa \citep{Takaya2001}. The perturbation streamfunction is $\psi' = g Z' / f(\varphi)$, where $Z' = Z_\mathrm{pert} - Z_\mathrm{base}$ is the geopotential height anomaly (m) and $f$ is the Coriolis parameter. The horizontal flux components are:

\begin{linenomath*}
\begin{equation}
  \mathbf{W} = \frac{p\cos\varphi}{2|\bar{\mathbf{U}}|a^2}
  \begin{pmatrix}
    \dfrac{\bar{u}}{\cos^2\!\varphi}\!\left[
      \left(\tfrac{\partial\psi'}{\partial\lambda}\right)^{\!2}
      - \psi'\tfrac{\partial^2\psi'}{\partial\lambda^2}
    \right]
    + \dfrac{\bar{v}}{\cos\varphi}\!\left[
      \tfrac{\partial\psi'}{\partial\lambda}\tfrac{\partial\psi'}{\partial\varphi}
      - \psi'\tfrac{\partial^2\psi'}{\partial\lambda\partial\varphi}
    \right]
    \\[6pt]
    \dfrac{\bar{u}}{\cos\varphi}\!\left[
      \tfrac{\partial\psi'}{\partial\lambda}\tfrac{\partial\psi'}{\partial\varphi}
      - \psi'\tfrac{\partial^2\psi'}{\partial\lambda\partial\varphi}
    \right]
    + \bar{v}\!\left[
      \left(\tfrac{\partial\psi'}{\partial\varphi}\right)^{\!2}
      - \psi'\tfrac{\partial^2\psi'}{\partial\varphi^2}
    \right]
  \end{pmatrix},
  \label{eq:tnflux}
\end{equation}
\end{linenomath*}

\noindent where $p$ is pressure, $a = 6.371 \times 10^6$~m is Earth's radius, and $(\bar{u}, \bar{v})$ are background horizontal winds clamped to $|\bar{\mathbf{U}}| \geq 5$~m~s$^{-1}$ to avoid singularities. All derivatives use centred finite differences on the full Aurora grid.

\subsection{Experimental Design}
\label{sec:expdesign}

Table~\ref{tab:exps} summarizes all experiments. A canonical configuration is fixed in Exp~1: forward FTLE at 500~hPa, $\tau=+48$~h, RK4 advection, 85th-percentile threshold, 500--1500~km TC distance; seeding at 925/850/700~hPa, $\eta=60\%$, $r=300$~km, $f_\mathrm{fall}=80\%$, min~RH~$=75\%$. Each subsequent experiment isolates one variable (Exp 2-7). Pacific Experiments~8--10 use Gaussian perturbations ($\Delta T=2.49$~K, $\sigma=8^\circ$) except Exp~10e (cloud-seeding scale). All forecasts are initialized from ERA5 at 2012-10-23~00Z; track deviation is reported as the Haversine distance between perturbed and baseline TC positions at $t = +168$~h.

\begin{table}[htbp]
\caption{Experimental design and track deviations at $t=+168$~h. Unlisted parameters match Exp~1 (Section~\ref{sec:expdesign}). Exps~1--7 use cloud-seeding physics ($\eta=60\%$, $r=300$~km, 925/850/700~hPa); Exps~8--10 use Gaussian perturbation ($\Delta T=2.49$~K, $\sigma=8^\circ$, 700--925~hPa) except Exp~10e.}
\label{tab:exps}
\centering\scriptsize
\setlength{\tabcolsep}{3pt}
\resizebox{\textwidth}{!}{%
\begin{tabular}{clp{4.4cm}p{4.2cm}r}
\toprule
Exp & Label & Parameter change vs Exp~1 & Site locations & Dev.~(km) \\
\midrule
\textbf{1} & \textbf{Local FTLE}
  & \textbf{Canonical — none}
  & \textbf{5.5$^\circ$N/82.3$^\circ$W,\ 8.75$^\circ$N/81.5$^\circ$W,\ 14.5$^\circ$N/90.8$^\circ$W}
  & \textbf{360.7} \\
\midrule
\multicolumn{5}{l}{\textit{Exp~2 — Sensitivity (FTLE and sites as Exp~1; seeding parameters varied)}} \\
2 & Radius        & $r = 100$~km                   & same as Exp~1 &  18.2 \\
  &               & $r = 10$~km                    & same as Exp~1 &   0.0 \\
  & Efficiency    & $\eta = 30\%$                  & same as Exp~1 & 120.8 \\
  &               & $\eta = 10\%$ (near-realistic) & same as Exp~1 &   9.1 \\
  & Levels        & 700/500/300~hPa seeding        & same as Exp~1 & 322.6 \\
  &               & 700~hPa seeding only           & same as Exp~1 & 302.7 \\
  & Single site   & Site~3 only                   & 14.5$^\circ$N/90.8$^\circ$W & 205.8 \\
\midrule
\multicolumn{5}{l}{\textit{Exp~3 — Random control (no FTLE; seeding as Exp~1)}} \\
3 & Random~1 (seed~42)   & No FTLE; random Caribbean sites & 12$^\circ$N/84$^\circ$W,\ 9$^\circ$N/71$^\circ$W,\ 25$^\circ$N/84$^\circ$W &  86.2 \\
  & Random~2 (seed~123)  &                                 & 23$^\circ$N/75$^\circ$W,\ 3$^\circ$N/71$^\circ$W,\ 15$^\circ$N/91$^\circ$W$^\dagger$ & 135.1 \\
  & Random~3 (seed~456)  &                                 & 5$^\circ$N/72$^\circ$W,\ 14$^\circ$N/72$^\circ$W,\ 1$^\circ$N/78$^\circ$W & 107.7 \\
  & Random~4 (seed~789)  &                                 & 13$^\circ$N/70$^\circ$W,\ 4$^\circ$N/80$^\circ$W,\ 8$^\circ$N/72$^\circ$W &  29.8 \\
  & Random~5 (seed~1234) &                                 & 17$^\circ$N/69$^\circ$W,\ 12$^\circ$N/92$^\circ$W,\ 14$^\circ$N/68$^\circ$W & 129.2 \\
  & \textit{Mean $\pm$ std} & & & \textit{97.6 $\pm$ 38.0} \\
\midrule
\multicolumn{5}{l}{\textit{Exp~4 — Backward FTLE control (seeding as Exp~1)}} \\
4 & Backward FTLE & $\tau = -48$~h; attracting LCS; subtropical ridge
  & 23.0$^\circ$N/75.3$^\circ$W,\ 23.0$^\circ$N/78.3$^\circ$W,\ 22.5$^\circ$N/72.3$^\circ$W & 91.5 \\
\midrule
\multicolumn{5}{l}{\textit{Exp~5 — Vertical-level control (seeding as Exp~1)}} \\
5 & 700~hPa FTLE & FTLE at 700~hPa; TC warm-core contamination
  & 3.0$^\circ$N/82.8$^\circ$W,\ 17.0$^\circ$N/66.5$^\circ$W,\ 4.5$^\circ$N/75.3$^\circ$W & 55.0 \\
\midrule
\multicolumn{5}{l}{\textit{Exp~6 — Upstream mid-latitude FTLE (seeding as Exp~1)}} \\
6 & Upstream FTLE & Expanded domain (0--60$^\circ$N); jet-ridge sites
  & 38.8$^\circ$N/50.8$^\circ$W,\ 36.3$^\circ$N/53.5$^\circ$W,\ 34.5$^\circ$N/56.5$^\circ$W & 26.8 \\
\midrule
\multicolumn{5}{l}{\textit{Exp~7 — Pacific delayed perturbation (cloud-seeding; FTLE recomputed at perturbation hour, Pacific domain)}} \\
7 & Pacific lead-3d & Perturb at $+$96~h; $\tau=48$~h
  & 52.8$^\circ$N/212.2$^\circ$E,\ 54.8$^\circ$N/151.2$^\circ$E,\ 49.8$^\circ$N/147.7$^\circ$E & 0.0 \\
\midrule
\multicolumn{5}{l}{\textit{Exp~8 — Pacific site survey (Gaussian $\Delta T=2.49$~K, $\sigma=8^\circ$, 700--925~hPa; $t=0$; 25 sites 30--50$^\circ$N/140--250$^\circ$E)}} \\
8 & Pacific best site    & Post-hoc best from grid survey (35$^\circ$N/165$^\circ$W) & 35$^\circ$N/195$^\circ$E  & 324.6 \\
8b & FTLE-guided Pacific  & Forward FTLE 500~hPa; 85th-pct sites in Pacific jet & 44$^\circ$N/140.5$^\circ$E,\ 43$^\circ$N/170.5$^\circ$E,\ 48$^\circ$N/144$^\circ$E & 99.9 \\
  & \textit{25-site mean} & Mean across all 25 grid sites & & \textit{167.6} \\
\midrule
\multicolumn{5}{l}{\textit{Exp~9/10 — Wave-regime Pacific perturbations (Gaussian as Exp~8; $t=0$)}} \\
9  & RWS-guided (top-3)         & Top-3 $|\mathrm{RWS}|$ at 200~hPa; includes dead-zone 45$^\circ$N/140$^\circ$E
   & 35$^\circ$N/195$^\circ$E,\ 45$^\circ$N/140$^\circ$E,\ 45$^\circ$N/167.5$^\circ$E & 136.0 \\
\textbf{10}  & \textbf{165$^\circ$W corridor (Gaussian)} & \textbf{All 3 productive corridor sites; Gaussian $\Delta T=2.49$~K, $\sigma=8^\circ$}
   & \textbf{30$^\circ$N/195$^\circ$E,\ 35$^\circ$N/195$^\circ$E,\ 40$^\circ$N/195$^\circ$E} & \textbf{508.1} \\
10e & 165$^\circ$W corridor (cloud-seeding) & Same 3 corridor sites; cloud-seeding ($\eta=60\%$, $r=300$~km, 925/850/700~hPa)
   & 30$^\circ$N/195$^\circ$E,\ 35$^\circ$N/195$^\circ$E,\ 40$^\circ$N/195$^\circ$E & 578.6 \\
\bottomrule
\multicolumn{5}{l}{$^\dagger$Random~2 Site~3 (15$^\circ$N/91$^\circ$W) lies within 0.75$^\circ$ of FTLE Site~3 (14.5$^\circ$N/90.75$^\circ$W), an accidental rediscovery of the optimal site.} \\
\multicolumn{5}{l}{All deviations at $t=+168$~h. Pacific Exps~8--10 use Gaussian perturbation except Exp~10e (cloud-seeding scale).}
\end{tabular}}
\end{table}

\section{Results}

\subsection{FTLE Field at Initialization}

Figure~\ref{fig:ftle_field} shows the 48-hour forward FTLE at 500~hPa on 2012-10-23~00Z across both perturbation domains. In the Caribbean, 89\% of cells are positive, with strong repelling ridges extending from southern Caribbean trade-wind boundaries northward into the subtropical ridge. The three FTLE-selected seeding sites (Table~\ref{tab:sites}) lie on positive FTLE ridges ($+$1.47 to $+$1.62~day$^{-1}$), well above the 85th-percentile threshold ($+$0.79~day$^{-1}$). Sites~1 and~2 (southern and western Caribbean inflow) straddle the dominant low-level inflow arms of Sandy's steering flow; Site~3 (subtropical jet entrance region) lies at the boundary between the trade-wind belt and the subtropical ridge. In the Pacific, high FTLE values are widespread in the jet stream, but the FTLE-selected Pacific sites (Section~\ref{sec:pacific_survey}) lie far north and west of the productive 165$^\circ$W corridor, explaining why FTLE guidance fails in this regime.

\begin{table}[htbp]
\caption{Seeding sites for Caribbean (Exp~1) and Pacific corridor (Exps~10, 10e) experiments. Caribbean FTLE values are from the corrected 500~hPa forward-FTLE field (85th-pct threshold $+$0.79~day$^{-1}$). Pacific individual deviations are from the Exp~8 single-site Gaussian grid survey at $t=+168$~h. Distance measured from Sandy's initial position (12.6$^\circ$N, 78.4$^\circ$W) at initialization.}
\label{tab:sites}
\centering\small
\begin{tabular}{clcccc}
\toprule
Site & Location & Dist. (km) & FTLE / Dev. & Role \\
\midrule
\multicolumn{5}{l}{\textit{Caribbean FTLE sites — Exp~1 (cloud-seeding, 925/850/700~hPa)}} \\
\midrule
1 & 5.50$^\circ$N, 82.25$^\circ$W  &   895 & $+$1.62~day$^{-1}$ & Southern Caribbean inflow arm \\
2 & 8.75$^\circ$N, 81.50$^\circ$W  &   591 & $+$1.60~day$^{-1}$ & Western Caribbean mid-level steering \\
3 & 14.25$^\circ$N, 91.00$^\circ$W & 1{,}375 & $+$1.47~day$^{-1}$ & Subtropical jet entrance region \\
\midrule
\multicolumn{5}{l}{\textit{Pacific corridor sites — Exps~10, 10e (165$^\circ$W, 30--40$^\circ$N); Gaussian or cloud-seeding}} \\
\midrule
1 & 30$^\circ$N, 165$^\circ$W (195$^\circ$E) & 8{,}989 & 263.1~km (indiv.) & Southern corridor, subtropical jet \\
2 & 35$^\circ$N, 165$^\circ$W (195$^\circ$E) & 8{,}903 & 324.6~km (indiv.) & Best single Pacific site \\
3 & 40$^\circ$N, 165$^\circ$W (195$^\circ$E) & 8{,}825 & 188.2~km (indiv.) & Northern corridor, jet core \\
\bottomrule
\multicolumn{5}{l}{Individual deviations (P1--P3) from single-site Gaussian runs in Exp~8 at $t=+168$~h.} \\
\multicolumn{5}{l}{Combined corridor (all 3 sites): Exp~10 Gaussian = 508.1~km; Exp~10e cloud-seeding = 578.6~km.}
\end{tabular}
\end{table}

\subsection{Caribbean FTLE Regime}

\subsubsection{Main Result and Atmospheric Evolution}

Table~\ref{tab:perturbation} documents the perturbation magnitudes at the three FTLE-selected sites. High moisture content in the lower troposphere produces large temperature anomalies: $+$21.0~K at 925~hPa, $+$17.0~K at 850~hPa, and $+$12.7~K at 700~hPa.

\begin{table}[htbp]
\caption{Perturbation magnitudes for the two canonical configurations. \textit{Top}: Exp~1 cloud-seeding at 925/850/700~hPa ($\eta=60\%$, $r=300$~km), averaged over the three Caribbean sites. \textit{Bottom}: Exps~10/10e at the 165$^\circ$W corridor; the Gaussian applies a uniform $\Delta T_0=2.49$~K regardless of RH, while the cloud-seeding variant (Exp~10e) uses the same ice-nucleation physics as Exp~1 with large local $\Delta T$ (7--14~K) but sparse vertical coverage where RH~$<$~75\%.}
\label{tab:perturbation}
\centering\small
\begin{tabular}{cccccc}
\toprule
Experiment & Level (hPa) & RH before (\%) & RH after (\%) & $\Delta T$ (K) & $\Delta q$ (g~kg$^{-1}$) \\
\midrule
\multicolumn{6}{l}{\textit{Exp~1 — Caribbean cloud-seeding ($\eta=60\%$, $r=300$~km, 925/850/700~hPa)}} \\
Exp~1 & 925 & $\sim$86 & $\sim$17 & $+$21.0 & $-$7.4 \\
      & 850 & $\sim$84 & $\sim$18 & $+$17.0 & $-$6.0 \\
      & 700 &      86  &      18  & $+$12.7 & $-$4.5 \\
\midrule
\multicolumn{6}{l}{\textit{Exp~10 — Pacific corridor Gaussian ($\Delta T_0=2.49$~K, $\sigma=8^\circ$, 700/850/925~hPa)}} \\
Exp~10 & 925/850/700 & N/A & N/A & $+$2.49 & 0 \\
\midrule
\multicolumn{6}{l}{\textit{Exp~10e — Pacific corridor cloud-seeding ($\eta=60\%$, $r=300$~km, 925/850/700~hPa)}} \\
Exp~10e & 925 & $\sim$99 & $\sim$20 & $+$14 & $-$4.6 \\
        & 850 & $\sim$97  & 20 & $+$12.6 & $-$3.9 \\
        & 700 & 96  & 20 & $+$7.4 & $-$2.5 \\
\bottomrule
\multicolumn{6}{l}{Exp~10e 925~hPa: all three corridor sites seeded. 850~hPa: only 30$^\circ$N and 40$^\circ$N seeded (35$^\circ$N RH=28\%).} \\
\multicolumn{6}{l}{700~hPa: only 40$^\circ$N seeded (30$^\circ$N RH=13\%, 35$^\circ$N RH=41\%). Local $\Delta T$ is larger than Gaussian's 2.49~K} \\
\multicolumn{6}{l}{per cell, but sparse vertical coverage limits coherent wave forcing relative to the synoptic Gaussian.}
\end{tabular}
\end{table}

\begin{figure}[p]
\centering
\noindent\includegraphics[width=\textwidth,height=0.88\textheight,keepaspectratio]{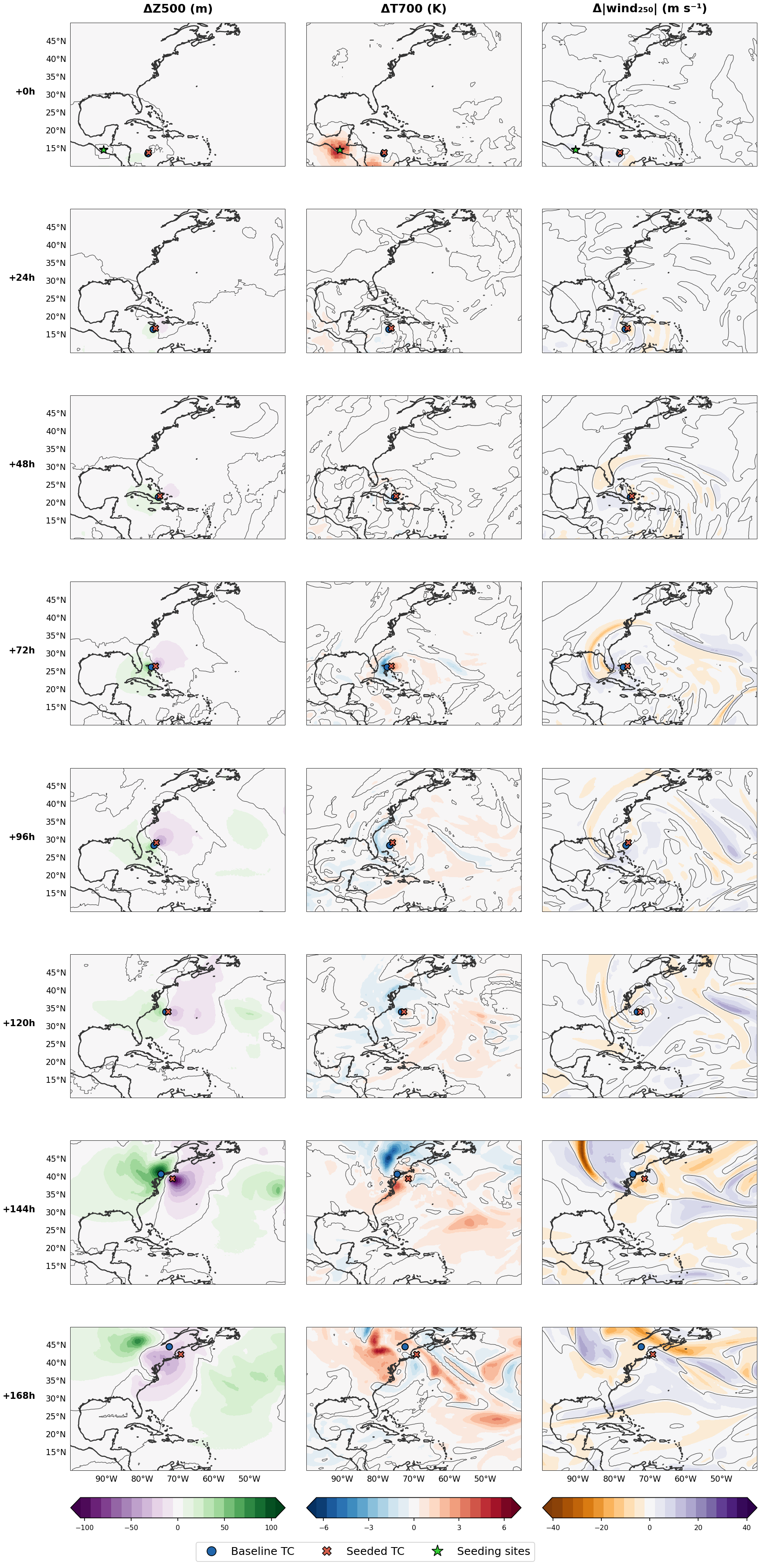}
\caption{Seeded--baseline atmospheric anomalies for Hurricane Sandy under Exp~1 (local FTLE-guided seeding). Three columns show (left) $\Delta Z_{500}$ (m), (center) $\Delta T_{700}$ (K), and (right) $\Delta|\mathrm{wind}_{250}|$ (m~s$^{-1}$), each at eight time steps from $+$0~h through $+$168~h (rows top to bottom). The perturbation begins as a localized $\Delta T_{700}$ warm anomaly confined to the seeding sites; large-scale anomalies ($\Delta Z_{500} > 50$~m, $\Delta|\mathrm{wind}_{250}| > 20$~m~s$^{-1}$) develop only after $+$120~h as Sandy's two diverging tracks interact differently with the mid-latitude trough. Blue dot: baseline TC; red cross: seeded TC; green stars: seeding sites.}
\label{fig:local_evolution}
\end{figure}

Figure~\ref{fig:local_evolution} shows the evolution of seeded--baseline atmospheric anomalies. The perturbation initially manifests as a localized warm anomaly at 700~hPa confined to the seeding sites, with negligible large-scale response. By $+$72~h, weak $\Delta Z_{500}$ anomalies appear over the western Atlantic. After $+$120~h, as Sandy recurves northward, both $\Delta Z_{500}$ and $\Delta|\mathrm{wind}_{250}|$ grow rapidly, reaching $\sim$50--80~m and $>$20~m~s$^{-1}$ by $+$156--168~h, a consequence of two diverging atmospheric states entering the mid-latitude baroclinic zone.

\subsubsection{Sensitivity to Perturbation Magnitude}

Track deviation exhibits strong nonlinear dependence on perturbation magnitude. Increasing the radius from 10~km (0~km deviation) to 100~km (18.2~km) to 300~km ($\sim$361~km) reveals a clear threshold behavior. Freeze efficiency shows a similarly steep response: $\eta = 10\%$ gives only 9.1~km, $\eta = 30\%$ gives 120.8~km, and $\eta = 60\%$ gives 360.7~km, with the main transition near $\eta \approx 20$--25\%. This strong nonlinearity reflects the recurvature-gate mechanism (Section~\ref{sec:mechanism}): perturbations that fail to deliver the critical $\sim$36~km positional offset by day~5 produce near-zero final deviations, while those that cross the threshold are catastrophically amplified.

Sensitivity to vertical structure shows that 700~hPa seeding alone recovers 302.7~km (84\% of canonical), confirming 700~hPa is the dominant contributor. The full 925/850/700~hPa configuration (360.7~km) adds modest improvement from higher moisture content at lower levels. Site-sensitivity tests show that Site~3 alone yields 205.8~km (57\% of canonical), confirming that the subtropical-ridge-adjacent site is the most influential single site, though all three contribute constructively.

\subsubsection{Control Experiments}
\label{sec:controls}

For five random site draws with identical seeding physics (Exp~3), deviations range from 29.8 to 135.1~km with a mean of 97.6~km ($\pm$38.0~km), giving an FTLE advantage of 3.7$\times$. The highest random result (135.1~km, seed~123) results from its third site lying within 0.75$^\circ$ of FTLE Site~3, an inadvertent rediscovery of the optimal steering-flow boundary that itself validates the FTLE targeting principle.

Backward FTLE (Exp~4) places sites at 22--23$^\circ$N in the subtropical ridge, outside Sandy's active steering flow. The result (91.5~km) is indistinguishable from the random mean; the deviation oscillates and re-converges near zero at $+$84~h and $+$132~h, confirming that sites outside the inflow envelope cannot sustain the positional offset to the recurvature gate.

Computing FTLE at 700~hPa (Exp~5) rather than 500~hPa selects sites near Sandy's warm-core vortex (3--17$^\circ$N), producing 55.0~km, below the random mean. This confirms that 500~hPa is required to capture the environmental steering flow rather than TC-internal Lagrangian structure. Together the three controls show that the FTLE advantage requires both the correct temporal direction (forward) and vertical level (500~hPa).

The mid-latitude upstream experiment (Exp~6) yields only 26.8~km despite strong seeding ($\Delta T = +$8 to $+$14~K) at sites 3,300--4,000~km from Sandy. The teleconnection to Sandy's Caribbean steering environment is too weak and slow for the 7-day forecast window, confirming that spatial location within the flow governs the response more than perturbation energy.

\subsection{Comprehensive Comparison: Deviation Timeseries and All Tracks}

\begin{figure}[htbp]
\noindent\includegraphics[width=\textwidth]{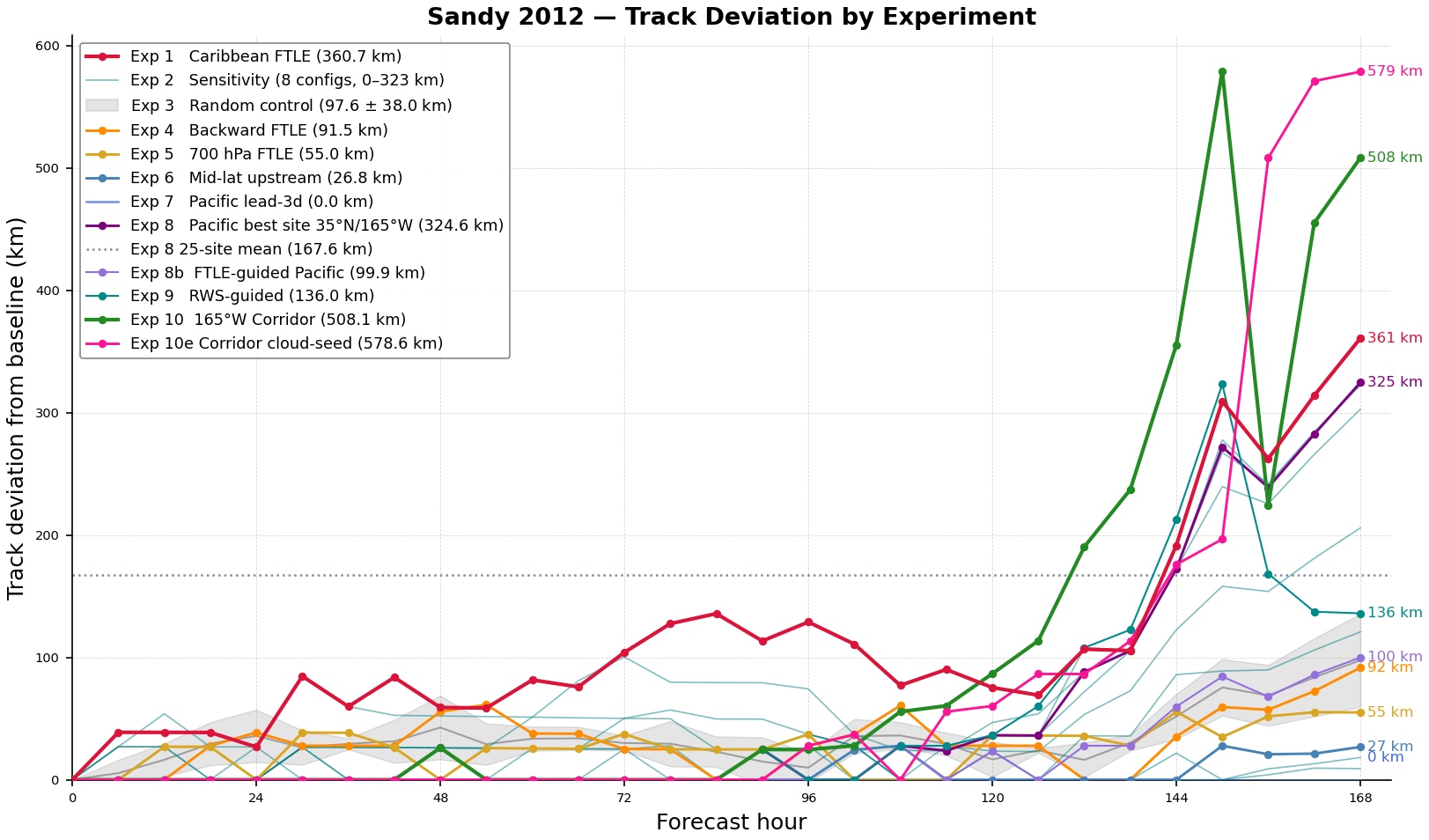}
\caption{Track deviation as a function of lead time for all major experiments.}
\label{fig:timeseries}
\end{figure}

Figure~\ref{fig:timeseries} shows the time evolution of track deviation for all major experiments. The FTLE-guided Caribbean perturbation (Exp~1, 360.7~km) and the Pacific corridor (Exp~10, 508.1~km) stand apart from all other experiments. Both exhibit the two-stage amplification pattern: a modest offset during days~1--5 followed by rapid growth as Sandy approaches the recurvature zone after $+$120~h. Control experiments (random mean, backward FTLE, 700~hPa FTLE) plateau well below the FTLE-guided result throughout the forecast.

\begin{figure}[htbp]
\noindent\includegraphics[width=\textwidth]{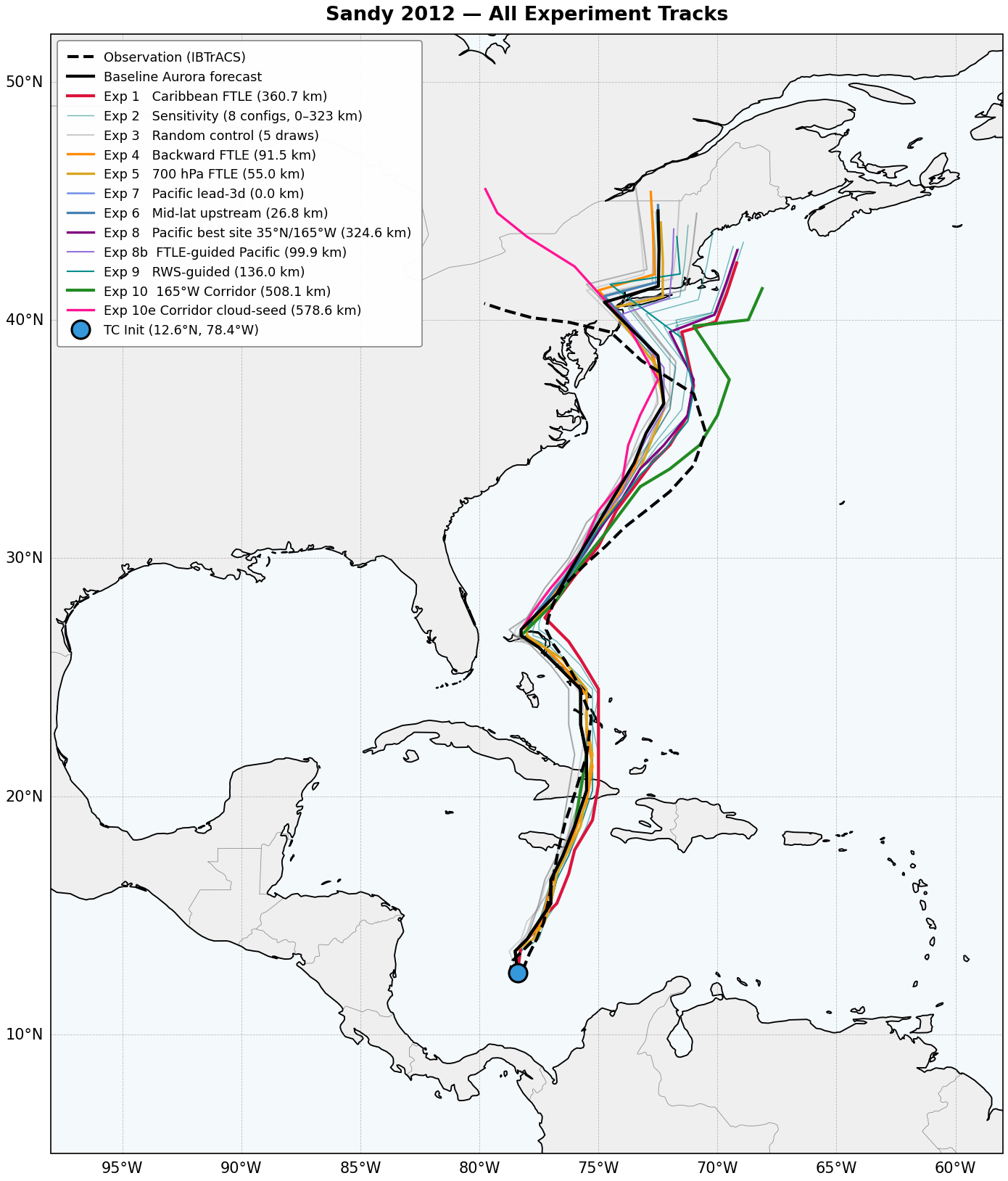}
\caption{TC track comparison for Hurricane Sandy (2012). Black dashed: IBTrACS best-track \citep{Knapp2010}. Black solid: Aurora baseline forecast. Red: Caribbean FTLE-guided (Exp~1, 360.7~km). Green: Pacific corridor (Exp~10, 508.1~km). The baseline captures Sandy's westward recurvature toward New Jersey. Both Exp 1 and Exp 10 perturbed tracks diverge southeastward, redirecting Sandy toward the open Atlantic.}
\label{fig:tracks}
\end{figure}

Figure~\ref{fig:tracks} shows the geographic TC tracks for the baseline, FTLE-seeded, Pacific corridor, and IBTrACS best-track. The baseline Aurora forecast captures Sandy's observed northward motion and westward recurvature toward New Jersey. The Caribbean FTLE-seeded track diverges southeastward from the baseline; the Pacific corridor track takes a similarly southeastward path, reaching open Atlantic rather than the New Jersey coast. The geographic divergence makes visible the physical separation between the two regimes: both ultimately redirect Sandy away from its historic landfall, but through distinct mechanisms operating in different parts of the globe.

\subsection{Pacific Rossby Wave Regime}
\label{sec:pacific_survey}

\subsubsection{25-Site Grid Survey}

A systematic 25-site grid survey (30--50$^\circ$N, 140--250$^\circ$E; Gaussian $\Delta T = 2.49$~K, $\sigma = 8^\circ$, 700--925~hPa) reveals a highly structured response (Table~\ref{tab:exps}). The most productive single site is 35$^\circ$N/195$^\circ$E ($\approx$165$^\circ$W), reaching 324.6~km; the 25-site mean is 167.6~km. Critically, all three sites at 195$^\circ$E longitude (30$^\circ$N: 263.1~km; 35$^\circ$N: 324.6~km; 40$^\circ$N: 188.2~km) substantially outperform sites at other longitudes, revealing a preferred Rossby wave corridor.

Forward FTLE-selected Pacific sites (44$^\circ$N/140.5$^\circ$E, 43$^\circ$N/170.5$^\circ$E, 48$^\circ$N/144$^\circ$E) produce only 99.9~k, below the 25-site mean, with Pearson $r = 0.317$ between FTLE values and site deviations. A Rossby wave source diagnostic at 200~hPa similarly fails ($r = 0.205$). These results establish that neither local Lagrangian instability nor instantaneous wave generation identifies the productive corridor; the governing factor is Rossby wave travel-time geometry.

\subsubsection{TN Wave Activity Flux Confirms the Rossby Wave Mechanism}

\begin{figure}[p]
\centering
\noindent\includegraphics[width=\textwidth,height=0.92\textheight,keepaspectratio]{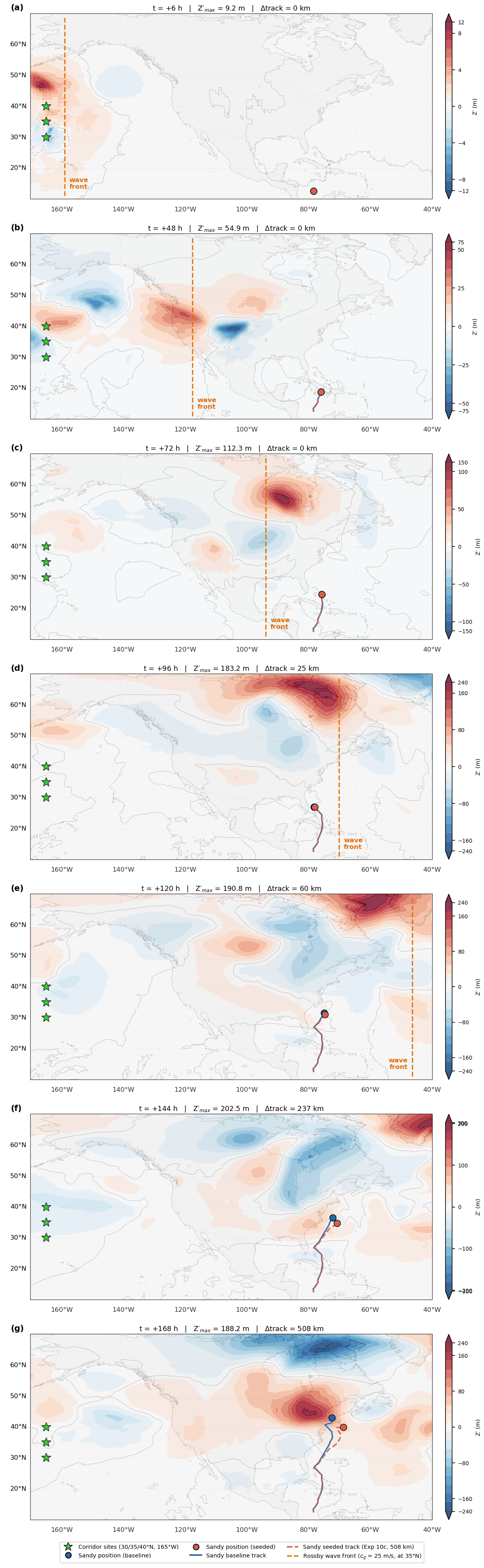}
\caption{Takaya-Nakamura wave activity flux at 250~hPa for the 165$^\circ$W corridor perturbation (Exp~10). Vectors show the horizontal flux $\mathbf{W}$ (m$^2$~s$^{-2}$ scaled by pressure); shading shows the perturbation geopotential height anomaly $Z'$ (m). Seven panels (a--g) at $t = +$6~h, $+$48~h, $+$72~h, $+$96~h, $+$120~h, $+$144~h, and $+$168~h. }
\label{fig:tn_flux}
\end{figure}

TN wave activity flux diagnostics computed from the Exp~10 corridor perturbation at 250~hPa confirm the Rossby wave propagation pathway (Figure~\ref{fig:tn_flux}). The perturbation height anomaly $Z'$ at 250~hPa grows from $\lesssim$6~m at $+$12~h to $\sim$124.5~m at $+$144~h; the mean TN flux magnitude $|\mathbf{W}|$ increases from $\sim$160~m$^2$~s$^{-2}$ at $+$6~h to $\sim$5$\times10^6$~m$^2$~s$^{-2}$ at $+$144~h, more than three orders of magnitude. The flux vectors propagate coherently eastward from the 195$^\circ$E source region, reaching the western Atlantic trough region by $+$96$-$120~h. This timing is consistent with Rossby group velocity $c_g \approx 25$~m~s$^{-1}$ propagating $\approx$8{,}000~km in $\approx$90~h, placing the wave energy arrival at Sandy's recurvature zone ($\approx$40$^\circ$N/65$^\circ$W) just as the storm enters the trough-interaction regime.

\subsubsection{Corridor Perturbation and Spatial Scale Sensitivity}

\begin{figure}[p]
\centering
\noindent\includegraphics[width=\textwidth,height=0.9\textheight,keepaspectratio]{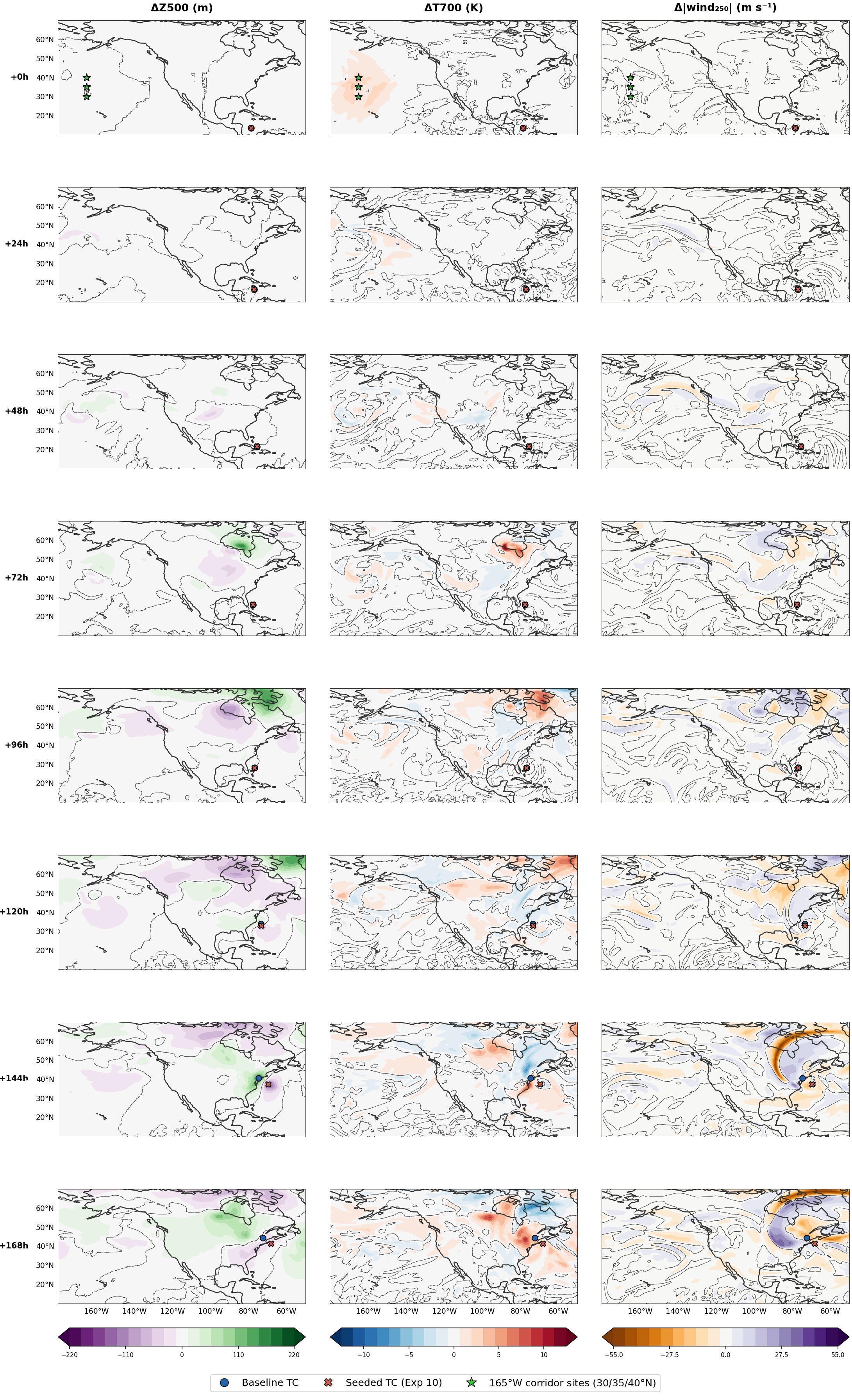}
\caption{Seeded--baseline atmospheric anomalies for the Pacific corridor perturbation (Exp~10). Three columns show (left) $\Delta Z_{500}$ (m), (center) $\Delta T_{700}$ (K), and (right) $\Delta|\mathrm{wind}_{250}|$ (m~s$^{-1}$), each at eight time steps from $+$0~h through $+$168~h (rows top to bottom). The perturbation begins as a $\Delta T_{700}$ warm anomaly at the Pacific corridor, other anomalies develop with wave east-ward progression. Blue dot: baseline TC; red cross: seeded TC; green stars: seeding sites.}
\label{fig:pacific_evolution}
\end{figure}

Applying the Gaussian perturbation simultaneously at all three corridor sites (30$^\circ$N, 35$^\circ$N, 40$^\circ$N/195$^\circ$E) yields 508.1~km (Exp~10), 57\% above the single best site (324.6~km) and 3.0$\times$ the 25-site mean (167.6~km). Figure~\ref{fig:pacific_evolution} shows the associated 500~hPa height anomaly evolution: a coherent Rossby wave train propagates eastward from the perturbation region, amplifying steadily and reaching the North Atlantic trough by $+$120$-$144~h. The three simultaneous source points create a broader north-south phase front that constructively reinforces the wave packet amplitude relative to a single source.

To test the role of perturbation spatial scale, Exp~10e applies the canonical cloud-seeding physics ($\eta = 60\%$, $r = 300$~km) at the same three corridor sites, replacing the synoptic Gaussian. ERA5 specific humidity at 925/850~hPa at these sites is 7--9~g~kg$^{-1}$ (RH~99\%), so where the RH threshold is met the local $\Delta T$ reaches 14~K per cell, larger than the Gaussian's 2.49~K. However, 700~hPa at 30$^\circ$N and 35$^\circ$N is too dry (RH~13--41\%) to qualify, and 850~hPa at 35$^\circ$N (RH~28\%) is also excluded, producing inconsistent vertical coverage across sites. This variant yields 578.6~km at $+$168~h (comparable to the Gaussian 508.1~km). The Gaussian's uniform 3D heating over $\sigma = 8^\circ \approx 890$~km excites the initial wave packet more efficiently; both converge to similar final deviations as the wave energy focuses on the recurvature gate.

\subsubsection{FTLE vs Corridor: Why FTLE Fails in the Pacific Regime}

The failure of Pacific FTLE guidance (99.9~km vs 508.1~km corridor) reflects a fundamental regime difference. In the Caribbean, Sandy's circulation creates LCS boundaries in the low-level steering flow; perturbations displace air parcels across these boundaries, producing a Lagrangian displacement that propagates with the steering flow. In the Pacific, perturbations must project onto Rossby waves that propagate via group velocity through the jet waveguide. The FTLE-selected Pacific sites (44--48$^\circ$N, western Pacific) identify regions of local Lagrangian instability, but these sites lie too far north and west for the wave packet to reach Sandy's recurvature zone within the 7-day forecast window.

By contrast, the RWS-guided top-3 sites (Exp~9) yield only 136.0~km --- below the single best site (324.6~km) --- because RWS guidance includes 45$^\circ$N/140$^\circ$E (0~km individually), a dead-zone site that causes destructive interference. Neither FTLE nor RWS identifies the corridor from $t=0$ local diagnostics; only geometric reasoning about Rossby wave propagation paths determines which sites are productive.

\section{Discussion}

\subsection{Two-Stage Amplification Mechanism}
\label{sec:mechanism}

Analysis of the forecast evolution reveals a two-stage amplification mechanism shared by both perturbation regimes. During days~1--5 ($+$0 to $+$120~h), the thermodynamic perturbation creates a persistent but modest positional offset of 36--50~km; the 250~hPa wind anomaly between seeded and baseline runs remains $\lesssim$0.02~m~s$^{-1}$, confirming no significant energy propagation to the jet stream during this phase.

The deviation timeline (Figure~\ref{fig:timeseries}) illustrates the two-stage mechanism. Stage~1 (0--120~h) shows modest 36--50~km positional offset from Caribbean seeding, or a delayed Rossby wave arrival from the Pacific corridor. Stage~2 ($>$120~h) shows bifurcation at the recurvature gate. Backward FTLE (91.5~km) reconverges to near-zero deviation twice ($+$84~h and $+$132~h), confirming that sites outside Sandy's direct steering flow cannot sustain the initial offset to the gate.

At approximately $+$120~h, Sandy reaches the recurvature zone where the interaction between the East Coast trough, the subtropical ridge, and the Greenland blocking anticyclone is highly sensitive to TC position \citep{Galarneau2013}. The 36~km offset at this moment places the seeded and baseline Sandy in qualitatively different trough-relative positions: the baseline is captured by the trough and turns West-northwest toward New Jersey; the seeded Sandy deflects southeastward. Between $+$120~h and $+$168~h the 250~hPa wind anomaly grows from $<$0.02 to 3.7~m~s$^{-1}$, and track deviation explodes from 36 to 361~km. This upper-level anomaly growth is a consequence of two diverging atmospheric states, not the cause of the track separation.

The non-monotonic deviation timeline (local minimum $\approx$37~km at $\approx$$+$96~h) occurs as both tracks briefly approach the same trough-relative position before the bifurcation locks in. The modest sensitivity to seeding pressure level (360.7~km vs 322.6~km; Table~\ref{tab:exps}) arises because any seeding configuration that delivers the critical 36~km positional offset by day~5 is sufficient to trigger the recurvature bifurcation; the exact seeding layer is immaterial once the threshold is crossed.

\subsection{FTLE Advantage and Spatial Specificity}

Forward FTLE guidance provides a 3.7$\times$ advantage over random placement through spatial specificity. The FTLE sites straddle Sandy's active 925--700~hPa inflow arms; the perturbation directly modifies the dominant inflow boundary rather than relying on downstream amplification. The absence of any upper-level response through day~5 (Section~3.2) confirms that the advantage arises from steering-flow placement, not upward energy propagation.

The sensitivity experiments reveal a threshold-controlled response consistent with minimum-energy control regimes identified in low-order Lorenz systems \citep{Liu2025_EGU}, where perturbations must cross a critical amplitude before the system bifurcates. The three control experiments (random: 97.6~km, backward FTLE: 91.5~km, 700~hPa FTLE: 55.0~km) collectively confirm that the advantage requires both the correct temporal direction (forward, not backward) and the correct vertical level (500~hPa, not 700~hPa). The backward FTLE selects sites at 22--23$^\circ$N in the subtropical ridge, outside Sandy's direct steering flow, and cannot sustain the initial offset. The 700~hPa FTLE selects sites within Sandy's warm-core vortex, directing energy toward TC-internal circulation boundaries rather than the environmental inflow arms. Only the combination of forward direction and environmental steering level (500~hPa) identifies the correct targeting region.

\subsection{Pacific Regime: Rossby Wave Teleconnection Mechanism}

The Pacific perturbation pathway operates through a fundamentally different physical mechanism. The Gaussian thermal anomaly at 925--700~hPa over the subtropical Pacific generates anomalous upper-level divergence that acts as a Rossby wave source \citep{Sardeshmukh1988}. The resulting wave packet propagates eastward along the October Pacific jet waveguide at $c_g \approx 25$~m~s$^{-1}$ \citep{HoskinsKaroly1981,HoskinsAmbrizzi1993}, reaching Sandy's recurvature zone ($\approx$40$^\circ$N/65$^\circ$W, $\approx$8{,}000~km from 165$^\circ$W) in $\approx$90~h. Sandy approaches this zone near $+$120~h, so a perturbation at $t = 0$ from the 165$^\circ$W corridor arrives with nearly optimal timing.

The stationary wavenumber theory of \citet{HoskinsAmbrizzi1993} predicts that the October Pacific jet acts as a waveguide for Rossby waves with specific zonal wavenumbers. The 165$^\circ$W longitude lies within the preferred excitation region for waves that propagate coherently to the western Atlantic, explaining the sharp longitudinal selectivity observed in the grid survey. Sites to the west ($<$195$^\circ$E) lie in a region of weaker jet or different wavenumber structure; sites to the east ($>$195$^\circ$E) have insufficient fetch for the wave packet to propagate before Sandy's recurvature window closes. This geometric constraint explains why neither FTLE ($r = 0.317$) nor RWS ($r = 0.205$) identifies the corridor from $t=0$ local diagnostics: both measure pointwise properties that do not capture the wave propagation path or resonance condition.

Sandy's extratropical transition \citep{Jones2003} makes the recurvature gate particularly sensitive to the upstream Rossby wave state \citep{Archambault2013}: a small modification to the trough amplitude or phase over the western Atlantic can determine whether Sandy turns westward into New Jersey or deflects southeastward into open ocean. The Pacific Rossby wave packet, arriving near $+$90$-$120~h, modifies the trough geometry that Sandy encounters at recurvature, producing the same qualitative bifurcation as the Caribbean perturbation but through a remote teleconnection rather than a local steering-flow modification.

\subsection{Two-Regime Synthesis}

The two perturbation regimes share a common amplification mechanism, the recurvature-gate bifurcation, but differ fundamentally in how they deliver the critical positional offset. In the Caribbean regime, forward FTLE at 500~hPa identifies repelling Lagrangian boundaries in Sandy's steering flow; mesoscale cloud-seeding ($r = 300$~km) at three FTLE-selected sites creates an offset that manifests within hours, yielding 360.7~km (3.7$\times$ the random mean). In the Pacific regime, the 165$^\circ$W corridor, identified by grid survey and confirmed by Takaya-Nakamura wave activity flux, acts through a Rossby wave teleconnection with an inherent, $\approx$90~h propagation delay; neither FTLE ($r = 0.317$) nor Rossby wave source diagnostics ($r = 0.205$) identify the corridor from $t = 0$ fields, because the operative constraint is wave travel-time geometry rather than local flow structure.

The Pacific regime achieves a larger final deviation (508.1~km Gaussian; 578.6~km cloud-seeding scale) than the Caribbean regime (360.7~km), but this comparison is not straightforward: the Gaussian perturbation ($\Delta T = 2.49$~K over $\sigma = 8^\circ$) represents a synoptic-scale forcing qualitatively different from the targeted mesoscale cloud-seeding in the Caribbean. The cloud-seeding corridor variant (Exp~10e: 578.6~km) demonstrates that localized thermodynamic forcing can also excite effective Rossby waves when placed at the 165$^\circ$W corridor, though the synoptic Gaussian is more efficient at wave excitation at pre-landfall timescales (237.2~km vs 113.4~km at $+$144~h). The fundamental distinction between regimes lies not in final deviation magnitude but in physical pathway: Lagrangian parcel steering versus Rossby wave teleconnection, each requiring a different diagnostic matched to its mechanism.

\subsection{Physical Realism and Limitations}
\label{sec:gap}

At $\eta = 60\%$ over $r = 300$~km radius, our cloud-seeding perturbation exceeds current operational capability: operational AgI campaigns achieve 5--15\% precipitation enhancement over 1{,}000--10{,}000~km$^2$ \citep{WMO2018,Cooper1997}. At near-realistic $\eta = 10\%$, the track deviation collapses to 9.1~km, confirming that the nonlinear threshold ($\eta \approx 20$--25\%) lies well above real-world seeding efficiency. These results establish the theoretical sensitivity upper bound and validate the FTLE targeting principle \citep{Liu2026_AR}, but they do not constitute a proposal for operational modification.

Sandy (2012) falls within Aurora's training period (ERA5 1979--2020); results are framed as a methodology demonstration on an in-sample case. The single-impulse perturbation design applied here does not reflect how an iterative campaign would operate in practice: repeated FTLE-guided seeding at refreshed sites over days~1--3 could potentially compound small positional offsets without requiring the large single-impulse magnitude demonstrated here. Evaluating such iterative strategies represents a natural extension of this work.

\section{Conclusions}

This study investigates the sensitivity of tropical cyclone trajectories to targeted atmospheric perturbations using Hurricane Sandy (2012) in the Aurora AI weather model. Two distinct perturbation pathways emerge. In the Caribbean, forward finite-time Lyapunov exponent (FTLE) diagnostics identify sensitive regions within Sandy's steering flow where perturbations produce substantially larger responses than random placement. In the Pacific, a preferred corridor near $165^\circ$W influences Sandy through Rossby-wave teleconnections. Although the physical mechanisms differ, both pathways generate large downstream track deviations by altering the large-scale circulation that governs storm motion.

A key finding is that both regimes share a common amplification mechanism. Initial perturbations produce only modest trajectory differences during the first several forecast days, but these differences are rapidly amplified when Sandy reaches the highly sensitive recurvature region associated with tropical--extratropical interaction. The largest track changes therefore arise not from direct manipulation of the cyclone itself, but from exploiting naturally occurring instabilities in the surrounding atmospheric flow.

The perturbation magnitudes required in these experiments remain well beyond current operational cloud-seeding capabilities, and the results should therefore be interpreted as a theoretical sensitivity analysis rather than a practical weather-modification strategy. More broadly, this work demonstrates how AI weather models can serve as computational laboratories for exploring atmospheric sensitivity, controllability, and intervention pathways in complex Earth systems. Future work should examine iterative low-amplitude perturbations, ensemble-based targeting strategies, and applications to a broader range of weather extremes.

\section*{Data Availability}

\begin{sloppypar}
ERA5 reanalysis data are available from the Copernicus Climate Data Store (\url{https://cds.climate.copernicus.eu}; \citealt{ECMWF2019}), and IBTrACS best-track data are available from NOAA/NCEI (\url{https://www.ncdc.noaa.gov/ibtracs}; \citealt{Knapp2010}). The Aurora model code is open source (\url{https://github.com/microsoft/aurora}; \citealt{Bodnar2025}). All analysis code will be made publicly available on GitHub upon acceptance.
\end{sloppypar}

\section*{Author Contributions}
Q.H. conceived the study, developed the code, ran all experiments, and wrote the manuscript. M.L. contributed methodology from the parallel AR perturbation study and reviewed the manuscript. Y.K. conducted Pacific upstream perturbation experiments and reviewed the manuscript. U.L. supervised the study and reviewed the manuscript.

\section*{Acknowledgments}
The authors thank Arizona State University for providing access to the SOL high-performance computing environment. This study received no external funding.

\bibliography{reference}

\end{document}